# On isotropic metric of Schwarzschild solution of Einstein equation

T. Mei

(Central China Normal University, Wuhan, Hubei PRO, P. R. of China
E-Mail: meitaowh@public.wh.hb.cn; meitao@mail.ccnu.edu.cn )

**Abstract** The static isotropic metric of Schwarzschild solution of Einstein equation cannot cover with the range of $r < \frac{2MG}{c^2}$, a new isotropic metric of Schwarzschild solution is obtained. The new isotropic metric has the characters: (1) It is dynamic and periodic. (2) It has infinite singularities of the spacetime. (3) It cannot cover with the range of $0 < r < r_0$; On the other hand, $r_0$ can be small discretionarily. (4) It seemed as if the range of negative $r$ could be unavoidable, although this range is meaningless for the Schwarzschild metric.

## 1 Presentment of the question

As well known, the Schwarzschild metric

$$\mathrm{d}s^2 = -\left(1 - \frac{2MG}{c^2 r}\right)c^2\mathrm{d}t^2 + \frac{1}{1 - \frac{2MG}{c^2 r}}\mathrm{d}r^2 + r^2\mathrm{d}\theta^2 + r^2\sin^2\theta\mathrm{d}\varphi^2 \tag{1}$$

can be changed to the static isotropic metric[1]

$$\mathrm{d}s^2 = -\left(\frac{1 - \frac{MG}{2c^2\rho}}{1 + \frac{MG}{2c^2\rho}}\right)^2 c^2\mathrm{d}t^2 + \left(1 + \frac{MG}{2c^2\rho}\right)^4 (\mathrm{d}\rho^2 + \rho^2\mathrm{d}\theta^2 + \rho^2\sin^2\theta\mathrm{d}\varphi^2) \tag{2}$$

by the coordinate transform

$$r = \rho\left(1 + \frac{MG}{2c^2\rho}\right)^2 . \tag{3}$$

However, if we write the above transform into the following form

$$\rho = \frac{\left(r - \frac{MG}{c^2}\right) \pm \sqrt{r\left(r - \frac{2MG}{c^2}\right)}}{2},$$

we can see obviously that the transform (3) holds only when $r \geq \frac{2MG}{c^2}$.

On the other hand, we can prove directly the above conclusion from (3). According to the basic inequality $a + b \geq 2\sqrt{ab}$ ($a$>0, $b$>0), we have

$$r = \left(\sqrt{\rho} + \frac{MG}{2c^2\sqrt{\rho}}\right)^2 \geq \left(2\sqrt{\sqrt{\rho}\cdot\frac{MG}{2c^2\sqrt{\rho}}}\right)^2 = \frac{2MG}{c^2}.$$

This conclusion means that the static isotropic metric expressed by (2) cannot cover with the range of $r < \frac{2MG}{c^2}$.

In this paper, we try to seek a coordinate transform

$$\left.\begin{aligned} r &= r(\rho,\sigma) \equiv \rho R(\rho,\sigma), \\ t &= t(\rho,\sigma), \end{aligned}\right\} \tag{4}$$

this transform can change the Schwarzschild metric expressed by (1) into the following form of isotropic metric:

$$\mathrm{d}s^2 = -W^2(\rho,\sigma)\mathrm{d}\sigma^2 + R^2(\rho,\sigma)(\mathrm{d}\rho^2 + \rho^2\mathrm{d}\theta^2 + \rho^2\sin^2\theta\mathrm{d}\varphi^2)\,. \tag{5}$$

For this purpose, inserting (4) into (1) and comparing with (5), we obtain the equations that the coordinate transform (4) must satisfy:

$$\left.\begin{aligned} &\left(1-\frac{2MG}{c^2 r}\right)c^2(t'_\sigma)^2 - \frac{1}{1-\frac{2MG}{c^2 r}}(r'_\sigma)^2 = W^2(\rho,\sigma), \\ &\left(1-\frac{2MG}{c^2 r}\right)c^2(t'_\rho)^2 - \frac{1}{1-\frac{2MG}{c^2 r}}(r'_\rho)^2 = -R^2(\rho,\sigma), \\ &\left(1-\frac{2MG}{c^2 r}\right)c^2 t'_\rho t'_\sigma - \frac{1}{1-\frac{2MG}{c^2 r}} r'_\rho r'_\sigma = 0. \end{aligned}\right\} \tag{6}$$

We can prove easily that, if (6) holds whatever $r < \frac{2MG}{c^2}$ or $r \geq \frac{2MG}{c^2}$, then $t'_\rho \neq 0, t'_\sigma \neq 0, r'_\rho \neq 0,$ and $r'_\sigma \neq 0$. This means that what we shall obtain is a dynamic metric.

## 2 The process of the calculation

For the sake of brevity, we set:

$$\frac{r}{2MG/c^2} \to r,\ \frac{t}{2MG/c^2} \to t,\ \frac{\rho}{2MG/c^2} \to \rho,\ \frac{\sigma}{2MG/c^2} \to \sigma, \tag{7}$$

then (6) is changed into the simpler form:

$$\left(1-\frac{1}{r}\right)c^2(t'_\sigma)^2 - \frac{1}{1-\frac{1}{r}}(r'_\sigma)^2 = W^2(\rho,\sigma), \tag{8}$$

$$\left(1-\frac{1}{r}\right)c^2(t'_\rho)^2 - \frac{1}{1-\frac{1}{r}}(r'_\rho)^2 = -R^2(\rho,\sigma), \tag{9}$$

$$\left(1-\frac{1}{r}\right)c^2 t'_\rho t'_\sigma - \frac{1}{1-\frac{1}{r}} r'_\rho r'_\sigma = 0. \tag{10}$$

In next process of calculation, we only list the main steps and results.

① $(r'_\rho r'_\sigma)\times(10)$:

$$t'_\rho t'_\sigma r'_\rho r'_\sigma = \frac{(r'_\rho)^2(r'_\sigma)^2}{c^2\left(1-\frac{1}{r}\right)^2}. \tag{11}$$

And then, (8)×(9) and using (10) and (11), we have

$$R^2W^2 = c^2(t'_\rho r'_\sigma - t'_\sigma r'_\rho)^2,$$

without lost universality, we set

$$t'_\sigma r'_\rho - t'_\rho r'_\sigma = \frac{RW}{c}. \tag{12}$$

Calculating $(r'_\sigma)^2 \times (9)$ and using (11) and (12)

$$(r'_\sigma)^2 R^2 = \frac{(r'_\rho)^2 (r'_\sigma)^2}{1-\frac{1}{r}} - \left(1-\frac{1}{r}\right) c^2 (t'_\rho r'_\sigma)^2 = c^2\left(1-\frac{1}{r}\right)\left(\frac{(r'_\rho)^2 (r'_\sigma)^2}{c^2\left(1-\frac{1}{r}\right)^2} - (t'_\rho r'_\sigma)^2\right)$$

$$= c^2\left(1-\frac{1}{r}\right)\left(t'_\rho t'_\sigma r'_\rho r'_\sigma - (t'_\rho r'_\sigma)^2\right) = c^2\left(1-\frac{1}{r}\right)\left(t'_\sigma r'_\rho - t'_\rho r'_\sigma\right) t'_\rho r'_\sigma = c^2\left(1-\frac{1}{r}\right)\frac{RW}{c} t'_\rho r'_\sigma,$$

we obtain:

$$t'_\rho = \frac{R}{Wc}\frac{r'_\sigma}{1-\frac{1}{r}}. \tag{13}$$

Inserting (13) into (10), we have

$$t'_\sigma = \frac{W}{Rc}\frac{r'_\rho}{1-\frac{1}{r}}. \tag{14}$$

Inserting (13) into (9), or inserting (14) into (8), and considering one of (4): $r = \rho R$, we obtain:

$$(r'_\rho)^2 - \frac{R^2}{W^2}(r'_\sigma)^2 = \left(1-\frac{1}{r}\right) R^2 = \left(1-\frac{1}{r}\right)\frac{r^2}{\rho^2} = \frac{r^2 - r}{\rho^2}. \tag{15}$$

Thus, the discussion for (8) ~ (10) is changed to that for (13) ~ (15).

② Calculating $(t'_\rho)'_\sigma$ by (13) and $(t'_\sigma)'_\rho$ by (14), respectively; And then, considering $(t'_\rho)'_\sigma = (t'_\sigma)'_\rho$, using (15) and $r = \rho R$, we obtain

$$r''_{\rho\rho} + \frac{R}{W}\left(\frac{W}{R}\right)'_\rho \cdot r'_\rho = \left(\frac{R}{W}\right)^2 r''_{\sigma\sigma} + \frac{R}{W}\left(\frac{R}{W}\right)'_\sigma \cdot r'_\sigma + \frac{1}{\rho^2}. \tag{16}$$

On the other hand, from (15) we have $\left(\frac{R}{W}\right)^2 = \frac{\rho^2 (r'_\rho)^2 - r^2 + r}{\rho^2 (r'_\sigma)^2}$ and $\left(\frac{W}{R}\right)^2 = \frac{\rho^2 (r'_\sigma)^2}{\rho^2 (r'_\rho)^2 - r^2 + r}$; according to these formulas, we calculate $\left(\frac{W}{R}\right)'_\rho$ and $\left(\frac{R}{W}\right)'_\sigma$. And then, inserting all the three expressions $\left(\frac{R}{W}\right)^2$, $\left(\frac{W}{R}\right)'_\rho$ and $\left(\frac{R}{W}\right)'_\sigma$ into (16), we obtain

$$2\rho^2 r r''_{\rho\rho} + 2\rho r r'_\rho - 4\rho^2 (r'_\rho)^2 + 2r^2 - 3r = 0.$$

This equation can be written to the following form:

$$\frac{\left(\rho^2 (r'_\rho)^2 - r^2 + r\right)'_\rho}{\rho^2 (r'_\rho)^2 - r^2 + r} = 4\frac{r'_\rho}{r}. \tag{17}$$

Calculating the integral of (17), we obtain:

$$\rho^2 (r'_\rho)^2 = C_0(\sigma) r^4 + r^2 - r. \tag{18}$$

$\rho^2 \times (15)$, considering $r = \rho R$ and using (18), we obtain

$$(r'_\sigma)^2 = C_0(\sigma) r^2 W^2, \tag{19}$$

(19) shown that $C_0(\sigma) \ge 0$. If $C_0(\sigma) = 0$, then we obtain the static isotropic metric expressed by (2), we therefore set $C_0(\sigma) \equiv D^2(\sigma) > 0$. Without lost universality, from (19) we have

$$r'_\sigma = DrW. \tag{20}$$

Now (18) can be written the following form:

$$\begin{aligned}\rho^2 (r'_\rho)^2 &= D^2 r^4 + r^2 - r = r(r - r_0)\left[D^2 r^2 + D^2 r_0 r + \left(D^2 r_0^2 + 1\right)\right] \\ &= D^2 r(r - r_0)\left[\left(r + \frac{r_0}{2}\right)^2 + \left(\frac{3}{4} r_0^2 + \frac{1}{D^2}\right)\right],\end{aligned} \tag{21}$$

where $r_0$ is the unique real root of the cubic equation $D^2 r^3 + r - 1 = 0$:

$$r_0 = \frac{1}{D}\left(\sqrt[3]{\frac{\sqrt{D^2 + \frac{4}{27}} + D}{2}} - \sqrt[3]{\frac{\sqrt{D^2 + \frac{4}{27}} - D}{2}}\right). \tag{22}$$

For $r_0$ we can prove easily

$$0 < r_0 < 1, \text{ whatever } D(\sigma) > 0 \text{ or } D(\sigma) < 0;$$

$$\lim_{D\to 0} r_0 = 1, \quad \lim_{D\to\pm\infty} r_0 = 0; \quad \lim_{D\to 0} Dr_0 = 0, \quad \lim_{D\to\pm\infty} Dr_0 = \pm\infty. \tag{23}$$

The condition that the right expression of (21) $\geq 0$ asks that the range of $r$ is $[r_0, \infty)$, i.e., $r_0 \leq r < \infty$, this means that the dynamic metric what we shall obtain finally cannot cover with the range of $(0, r_0)$; On the other hand, $r_0$ can be small discretionarily if and only if $|D|$ is big enough.

And, also from (21), other allowable the range of $r$ is $-\infty < r < 0$. Of course, this range is meaningless for the Schwarzschild metric expressed by (1).

③ (21) belongs to the type of elliptic integral, according to the standard theory of elliptic functions[2,3] we can find out $\lambda_+, \lambda_-, A_+$ and $A_-$:

$$\sqrt{1 - \lambda_\pm} = \sqrt{3 + r_0^{-2} D^{-2}} \mp \sqrt{1 + r_0^{-2} D^{-2}}, \quad A_\pm = \frac{\sqrt{1 + r_0^{-2} D^{-2}}}{\sqrt{1 - \lambda_\pm}}, \tag{24}$$

we can prove easily

$$2\sqrt{3} - 3 < \lambda_+ < 1, \; -\infty < \lambda_- < -(2\sqrt{3} + 3); \; \frac{\sqrt{3}+1}{2} < A_+ < \infty, \; \frac{\sqrt{3}-1}{2} < A_- < \frac{1}{2}; 0 < \frac{A_-}{A_+} < 2 - \sqrt{3}. \tag{25}$$

$$\left.\begin{aligned} r(r - r_0) &= \frac{1 - \lambda_-}{\lambda_+ - \lambda_-}(r - r_0 A_-)^2 - \frac{1 - \lambda_+}{\lambda_+ - \lambda_-}(r + r_0 A_+)^2, \\ D^2 r^2 + D^2 r_0 r + \left(D^2 r_0^2 + 1\right) &= D^2 \lambda_+ \frac{1 - \lambda_-}{\lambda_+ - \lambda_-}(r - r_0 A_-)^2 - D^2 \lambda_- \frac{1 - \lambda_+}{\lambda_+ - \lambda_-}(r + r_0 A_+)^2. \end{aligned}\right\} \tag{26}$$

Based on (26), (21) can be written to the following form:

$$\begin{aligned}\rho^2 (r'_\rho)^2 &= D^2 r^4 + r^2 - r = r(r - r_0)\left[D^2 r^2 + D^2 r_0 r + \left(D^2 r_0^2 + 1\right)\right] \\ &= D^2 \lambda_+ \left(\frac{1 - \lambda_-}{\lambda_+ - \lambda_-}\right)^2 (r + r_0 A_+)^4 \left[\left(\frac{r - r_0 A_-}{r + r_0 A_+}\right)^2 - \left(\frac{A_-}{A_+}\right)^2\right]\left[\left(\frac{r - r_0 A_-}{r + r_0 A_+}\right)^2 + \left(\frac{\sqrt{-\lambda_-}}{\sqrt{\lambda_+}} \frac{A_-}{A_+}\right)^2\right].\end{aligned} \tag{27}$$

According to the standard theory of elliptic functions[2,3] we take the transformation of variable

$$u = \frac{r - r_0 A_-}{r + r_0 A_+}, \quad r = r_0 \frac{A_- + A_+ u}{1 - u}; \tag{28}$$

Note that

$$\frac{\sqrt{1 - \lambda_+}}{\sqrt{1 - \lambda_-}} = \frac{1 - A_-}{1 + A_+} = \frac{A_-}{A_+} \leq u \leq 1^-, \quad \text{for} \quad r_0 \leq r < \infty. \tag{29}$$

(27) is changed to the form

$$\rho^2 (u'_\rho)^2 = D^2 r_0^2 \lambda_+ \left(\frac{1-\lambda_-}{\lambda_+ - \lambda_-}\right)^2 (A_+ + A_-)^2 \left[u^2 - \left(\frac{A_-}{A_+}\right)^2\right]\left[u^2 + \left(\frac{\sqrt{-\lambda_-}}{\sqrt{\lambda_+}}\frac{A_-}{A_+}\right)^2\right]. \tag{30}$$

And, further, according to the standard theory of elliptic function[2,3] , we take the transformation of variable

$$u^2 = \frac{(A_- / A_+)^2}{1-v^2}, \tag{31}$$

(30) is changed to the standard Jacobian form

$$\rho^2 (v'_\rho)^2 = A^2 \left(1-v^2\right)\left(1-k^2 v^2\right), \tag{32}$$

where

$$\left.\begin{aligned} k &= \frac{\sqrt{-\lambda_-}}{\sqrt{\lambda_+ - \lambda_-}} = \frac{\sqrt{\left(\sqrt{3D^2 r_0^2+1}+\sqrt{D^2 r_0^2+1}\right)\left(\sqrt{3D^2 r_0^2+1}+3\sqrt{D^2 r_0^2+1}\right)}}{2\sqrt{2}\sqrt[4]{3D^2 r_0^2+1}\sqrt[4]{D^2 r_0^2+1}}, \\ A &= \sqrt{D^2 r_0^2 \lambda_+ \left(\frac{1-\lambda_-}{\lambda_+ - \lambda_-}\right)^2 (A_+ + A_-)^2 \left(\frac{A_-}{A_+}\right)^2 \left[1+\left(\frac{\sqrt{-\lambda_-}}{\sqrt{\lambda_+}}\right)^2\right]} \\ &= \sqrt[4]{3D^2 r_0^2+1}\sqrt[4]{D^2 r_0^2+1} = \frac{\sqrt[4]{3-2r_0}}{\sqrt{r_0}}, \end{aligned}\right\} \tag{33}$$

where the formula $D^2 r_0^2 = \frac{1-r_0}{r_0}$ is used, and we can prove $\frac{\sqrt{2+\sqrt{3}}}{2} < k < 1$. Now we set

$$\xi = \ln \rho + B(\sigma), \tag{34}$$

from (32) we therefore obtain

$$v = \mathrm{sn}(A\xi, k), \tag{35}$$

where $\mathrm{sn}(A\xi, k)$ is one of Jacobian elliptic functions.

From (31), $u = \pm \frac{A_-/A_+}{\sqrt{1-v^2}}$; according to (28) we can prove that $u = -\frac{A_-/A_+}{\sqrt{1-v^2}}$ corresponds to $-r_0 A_+ \le r \le 0$; and (29) shows that what we need is $\frac{A_-}{A_+} \le u \le 1^-$, namely, $r_0 \le r < \infty$; we therefore take

$$u = +\frac{A_-/A_+}{\sqrt{1-v^2}} = \frac{A_-/A_+}{\sqrt{1-\mathrm{sn}^2(A\xi,k)}} = \frac{A_-/A_+}{\mathrm{cn}(A\xi,k)}. \tag{36}$$

But we must note that the transformation of variable (31), and further, (36) leads to the change of the range of *r*, because $0 \le \mathrm{cn}(A\xi, k) \le 1$, the *u* given by (36) satisfies $\frac{A_-}{A_+} \le u < \infty$. We divide the area $\left[\frac{A_-}{A_+}, \infty\right)$ into two areas: $\left[\frac{A_-}{A_+}, 1^-\right]$ and $[1^+, \infty)$; according to (29), when $\frac{A_-}{A_+} \le u \le 1^-$, $r_0 \le r < \infty$, this is what we want to need; but when $1^+ \le u < \infty$, $-\infty < r < -r_0 A_+$. The important key is that, in this case, differing from known Lemaitre, Eddington-Finkelstein, Novikow and Kruskal coordinates[4~6], from (28) and (36) we see, it seemed as if the range of negative *r* could be unavoidable if we use the standard theory of elliptic function.

In spite of the question of the range of negative $r$ provisionally, from (28) and (36) we obtain

$$r(\rho,\sigma) = r_0 A_- \frac{\mathrm{cn}(A\xi,k)+1}{\mathrm{cn}(A\xi,k)-\dfrac{A_-}{A_+}}, \tag{37}$$

$$R(\rho,\sigma) = \frac{r(\rho,\sigma)}{\rho} = \frac{r_0 A_-}{\rho}\frac{\mathrm{cn}(A\xi,k)+1}{\mathrm{cn}(A\xi,k)-\dfrac{A_-}{A_+}}. \tag{38}$$

From (37) we see

$$\left.\begin{array}{lll} \left(\dfrac{A_-}{A_+}\right)^+ \le \mathrm{cn}(A\xi,k) \le 1 & \text{corresponds to} & r_0 \le r < \infty; \\ \mathrm{cn}(A\xi,k) = \dfrac{A_-}{A_+} & \text{corresponds to} & r = \infty; \\ 0 \le \mathrm{cn}(A\xi,k) \le \left(\dfrac{A_-}{A_+}\right)^- & \text{corresponds to} & -\infty < r \le -r_0 A_+ . \end{array}\right\} \tag{39}$$

It seemed as if "arriving" the range of negative $r$ is after "striding over" $r=\infty$.

Now that the range of negative $r$ is unavoidable, we should not give up indiscreetly the range of $-r_0 A_+ \le r \le 0$, namely,

$$u = -\frac{A_-/A_+}{\sqrt{1-v^2}} = -\frac{A_-/A_+}{\sqrt{1-\mathrm{sn}^2(A\xi,k)}} = -\frac{A_-/A_+}{\mathrm{cn}(A\xi,k)}, \tag{40}$$

from (28) and (40), for the case of $-r_0 A_+ \le r \le 0$,

$$r(\rho,\sigma) = r_0 A_- \frac{\mathrm{cn}(A\xi,k)-1}{\mathrm{cn}(A\xi,k)+\dfrac{A_-}{A_+}}, \tag{41}$$

$$R(\rho,\sigma) = \frac{r(\rho,\sigma)}{\rho} = \frac{r_0 A_-}{\rho}\frac{\mathrm{cn}(A\xi,k)-1}{\mathrm{cn}(A\xi,k)+\dfrac{A_-}{A_+}}. \tag{42}$$

Note that $\mathrm{cn}(A\xi,k)=1$ corresponds to $r$=0.

So far, $D=D(\sigma)$ is a function of the variable $\sigma$, now we set that $D$ is a constant, thus, all $r_0, \lambda_+, \lambda_-, A_+, A_-, k$ and $A$ are constants. From (20), (37) and (41) we have:

$$W(\rho,\sigma) = \frac{r'_\sigma}{Dr} \equiv W_0(\rho,\sigma)\frac{\mathrm{d}B(\sigma)}{\mathrm{d}\sigma}, \tag{43}$$

$$W_0(\rho,\sigma) = \frac{A^3}{D^3 r_0^2 A_+}\frac{\mathrm{sn}(A\xi,k)\cdot\mathrm{dn}(A\xi,k)}{\left(\mathrm{cn}(A\xi,k)+1\right)\left(\mathrm{cn}(A\xi,k)-\dfrac{A_-}{A_+}\right)}, \quad \text{for the case of } r_0 \le r < \infty \text{ and } -\infty < r \le -r_0 A_+; \tag{44}$$

or

$$W_0(\rho,\sigma) = -\frac{A^3}{D^3 r_0^2 A_+}\frac{\mathrm{sn}(A\xi,k)\cdot\mathrm{dn}(A\xi,k)}{\left(\mathrm{cn}(A\xi,k)-1\right)\left(\mathrm{cn}(A\xi,k)+\dfrac{A_-}{A_+}\right)}, \text{ for the case of } -r_0 A_+ \le r \le 0. \tag{45}$$

④ For seeking the coordinate transform of $t=t(\rho,\sigma)$, inserting (20) into (13), we obtain

$$t'_\rho = \frac{R}{c}\frac{Dr}{1-\dfrac{1}{r}} = \frac{D}{c}\frac{1}{\rho}\frac{r^3}{r-1}; \tag{46}$$

From (37) and (41) we have

$$r'_\rho=\frac{\mathrm{d}r}{\mathrm{d}\xi}\frac{\partial\xi}{\partial\rho}=\frac{1}{\rho}\frac{\mathrm{d}r}{\mathrm{d}\xi},\quad r'_\sigma=\frac{\mathrm{d}r}{\mathrm{d}\xi}\frac{\partial\xi}{\partial\sigma}=\frac{\mathrm{d}r}{\mathrm{d}\xi}\frac{\mathrm{d}B(\sigma)}{\mathrm{d}\sigma}, \tag{47}$$

comparing one of the above formulas, $r'_\rho=\frac{1}{\rho}\frac{\mathrm{d}r}{\mathrm{d}\xi}$, with (21), we have

$$\left(\frac{\mathrm{d}r}{\mathrm{d}\xi}\right)^2=D^2r^4+r^2-r. \tag{48}$$

Using (20), (47) and (48), (14) can be written to the form

$$t'_\sigma=\frac{\rho r'_\sigma}{Dr^2c}\frac{r'_\rho}{1-\frac{1}{r}}=\frac{1}{Dr^2c}\frac{1}{1-\frac{1}{r}}\left(\frac{\mathrm{d}r}{\mathrm{d}\xi}\right)^2\frac{\mathrm{d}B(\sigma)}{\mathrm{d}\sigma}=\frac{D}{c}\frac{r^3}{r-1}\frac{\mathrm{d}B(\sigma)}{\mathrm{d}\sigma}+\frac{1}{cD}\frac{\mathrm{d}B(\sigma)}{\mathrm{d}\sigma}. \tag{49}$$

If we define a function $T=T(r)$

$$T(r)=\frac{D}{c}\int^r\frac{r^3}{(r-1)\sqrt{D^2r^4+r^2-r}}\mathrm{d}r, \tag{50}$$

then we can prove that the transform

$$t=t(\rho,\sigma)=T(r)+\frac{B(\sigma)}{cD}=T(r(\rho,\sigma))+\frac{B(\sigma)}{cD} \tag{51}$$

satisfies (46) and (49) by calculating $t'_\rho=\frac{\mathrm{d}T(r)}{\mathrm{d}r}r'_\rho$ and $t'_\sigma=\frac{\mathrm{d}T(r)}{\mathrm{d}r}r'_\sigma+\frac{1}{cD}\frac{\mathrm{d}B(\sigma)}{\mathrm{d}\sigma}$, respectively.

## 3 Result and discussion

Inserting (38) and (40) into (5), and considering (7), we obtain a dynamic isotropic metric of Schwarzschild field

$$\mathrm{d}s^2=-W_0^2(\rho,\sigma)\left(\frac{\mathrm{d}B(\sigma/(2MG/c^2))}{\mathrm{d}(\sigma/(2MG/c^2))}\right)^2\mathrm{d}\sigma^2+\left(\frac{2MG}{c^2}\right)^2R^2(\rho,\sigma)(\mathrm{d}\rho^2+\rho^2\mathrm{d}\theta^2+\rho^2\sin^2\theta\mathrm{d}\varphi^2)$$

$$=-W_0^2(\rho,\sigma)\left(\frac{2MG}{c^2}\right)^2\left(\mathrm{d}B\left(\frac{\sigma}{2MG/c^2}\right)\right)^2+\left(\frac{2MG}{c^2}\right)^2R^2(\rho,\sigma)(\mathrm{d}\rho^2+\rho^2\mathrm{d}\theta^2+\rho^2\sin^2\theta\mathrm{d}\varphi^2),$$

$$\xi=\ln\frac{\rho}{2MG/c^2}+B\left(\frac{\sigma}{2MG/c^2}\right).$$

We can redefine the time scaling $B\left(\frac{\sigma}{2MG/c^2}\right)\to\pm\frac{\sigma}{2MG/c^2}$ and obtain the final form

$$\left.\begin{aligned}&\mathrm{d}s^2=-W_0^2(\rho,\sigma)\mathrm{d}^2\sigma+\left(\frac{2MG}{c^2}\right)^2R^2(\rho,\sigma)(\mathrm{d}\rho^2+\rho^2\mathrm{d}\theta^2+\rho^2\sin^2\theta\mathrm{d}\varphi^2),\\&\xi=\ln\frac{\rho}{2MG/c^2}\pm\frac{\sigma}{2MG/c^2}.\end{aligned}\right\} \tag{52}$$

where $R(\rho,\sigma)$ and $W_0(\rho,\sigma)$ are expressed by (38) and (44) for the case of $r_0\le r<\infty$ and $-\infty<r\le-r_0A_+$, or by (42) and (45) for the case of $-r_0A_+\le r\le 0$, respectively.

The dynamic isotropic metric expressed by (52) has the following characters:

① The metric is periodic, because all the periods of the Jacobian elliptic functions $\mathrm{sn}(A\xi,k)$, $\mathrm{cn}(A\xi,k)$ and $\mathrm{dn}(A\xi,k)$ are the same. The reason that the metric is periodic is obvious: the all periodic functions $\mathrm{sn}(A\xi,k)$, $\mathrm{cn}(A\xi,k)$ and $\mathrm{dn}(A\xi,k)$ are introduced by the coordinate transform (37) and (51), or (41) and (51), respectively.

② There are infinite singularities of the spacetime at $\mathrm{cn}(A\xi,k)=\frac{A_-}{A_+}$, from (39) we know that all these singularities correspond to $r=\infty$ for the case of $r_0 \le r < \infty$ and $-\infty < r \le -r_0 A_+$; and at $\mathrm{cn}(A\xi,k)=1$, from (41) we know that all these singularities correspond to $r$=0 for the case of $-r_0 A_+ \le r \le 0$. Another singularity is $\rho = 0$ for the both cases. However, from

$$
R_{\mu\nu\rho\sigma}R^{\mu\nu\rho\sigma} = \left(\frac{2MG}{c^2}\right)^2 \frac{12}{r^6}
$$

$$
= \begin{cases} \left(\dfrac{c^2}{2MG}\right)^4 \dfrac{12}{(r_0 A_-)^6} \dfrac{\left(\mathrm{cn}(A\xi,k) - \dfrac{A_-}{A_+}\right)^6}{(\mathrm{cn}(A\xi,k)+1)^6}, & \text{for the case of } r_0 \le r < \infty \text{ and } -\infty < r \le -r_0 A_+; \\ \left(\dfrac{c^2}{2MG}\right)^4 \dfrac{12}{(r_0 A_-)^6} \dfrac{\left(\mathrm{cn}(A\xi,k) + \dfrac{A_-}{A_+}\right)^6}{(\mathrm{cn}(A\xi,k)-1)^6}, & \text{for the case of } -r_0 A_+ \le r \le 0 \end{cases} \tag{53}
$$

we can see obviously that the spacetime are well-behaved at all singularities satisfying $\mathrm{cn}(A\xi,k) = \dfrac{A_-}{A_+}$ for the case of $r_0 \le r < \infty$ and $-\infty < r \le -r_0 A_+$, and has infinite honest singularities when $\mathrm{cn}(A\xi,k)=1$ which correspond to $r$=0 for the case of $-r_0 A_+ \le r \le 0$, respectively; as for the point of $\rho = 0$, the spacetime is still well-behaved, although $R_{\mu\nu\rho\sigma}R^{\mu\nu\rho\sigma}$ is uncertain (because $\lim_{\rho\to 0} \ln\rho = -\infty$), at that point.

③ The metric cannot cover with the range of $(0, r_0)$, on the other hand, $r_0$ can be small discretionarily if and only if $|D|$ is big enough.

④ The application of the standard theory of elliptic functions leads us to have to face the range of negative $r$, even if we try to look for other transformation of variable but not that of (31), because the metric expressed by (52) is a solution of the Einstein equation $R_{\mu\nu} = 0$, we can also ask the question that what is the relation between the metric expressed by (52) and the Schwarzschild metric expressed by (1)?

How to understand this property? Maybe, is there "space outside space" just as there exists "time after time" (for example, a infalling astronaut to cross $r = \dfrac{2MG}{c^2}$ needs infinite time from our point of view, however, a finite fixed interval of his proper coordinates)?

A point of view deems that $r$ is as time coordinate when $r < \dfrac{2MG}{c^2}$, if so then the case of $r<0$ is comprehensible: It means the time before the moment of $r$=0. However, from (1) we can see that, formally, $r$ is still space coordinate when $r<0$.

This question will be studied in further.